\documentclass[preprint,showpacs,preprintnumbers,amsmath,amssymb]{revtex4}
\usepackage{epsfig}
\usepackage{graphicx}
\usepackage{dcolumn}
\usepackage{bm}

\let\jnfont=\rm
\def\NPB#1,{{\jnfont Nucl.\ Phys.\ B }{\bf #1},}
\def\PLB#1,{{\jnfont Phys.\ Lett.\ B }{\bf #1},}
\def\EPJC#1,{{\jnfont Eur.\ Phys.\ Jour.\ C }{\bf #1},}
\def\PRD#1,{{\jnfont Phys.\ Rev.\ D }{\bf #1},}
\def\PRL#1,{{\jnfont Phys.\ Rev.\ Lett.\ }{\bf #1},}
\def\MPLA#1,{{\jnfont Mod.\ Phys.\ Lett.\ A }{\bf #1},}
\def\JPG#1,{{\jnfont J.\ Phys.\ G}{\bf #1},}
\def\CTP#1,{{\jnfont Commun.\ Theor.\ Phys.\ }{\bf #1},}
\def\ZPC#1,{{\jnfont Z.\ Phys.\ C }{\bf #1},}
\def\RMP{\jnfont  Rev. Mod. Phys.}
\def\CPC#1,{{\jnfont Chin. \ Phys. \ C}{\bf #1},}
\def\CPL#1,{{\jnfont Chin. \ Phys. \ Lett}{\bf #1},}
\def\JHEP#1,{{\jnfont JHEP \ }{\bf #1},}
\def\q_slash{\not{\hbox{\kern-2.1pt $q$}}}
\def\p_slash{\not{\hbox{\kern-4.0pt $p$}}}
\def\k_slash{\not{\hbox{\kern-2.1pt $k$}}}
\def\E_slash{\not{\hbox{\kern-2.1pt $E$}}}

\begin{document}

\title{The charged top-pion production associated with the bottom quark pair as a probe of the
topcolor-assisted technicolor model at the LHC }
\author{Guo-Li Liu$^a$$^{,*}$\footnotetext{*Email address: guoliliu@zzu.edu.cn},
Ping Zhou$^{a,b}$$^{,\dagger}$\footnotetext{$^\dagger$Email address:
pzhou@fzd.de}}
\address{$^a$ Department of Physics, ZhengZhou University, Zhengzhou, Henan, China
\\ $^b$ Institut f\"{u}r Strahlenphysik, Forschungszentrum Dresden-Rossendorf, 01314~Dresden,
 Germany}

\begin{abstract}
The topcolor-assisted technicolor (TC2) model predicts the existence
of the charged top-pions ($\pi_t^\pm$), whose large couplings with
the third generation fermions will induce the charged top-pion
production associated with the bottom and anti-bottom quark pair at
the CERN Large Hadron Collider (LHC) through the parton processes
$c\bar b \to \pi_t^+ b \bar b$ and  $u\bar d(c\bar s) \to \pi_t^+ b
\bar b$. In this paper we examine these productions and find that,
due to the small Standard Model backgrounds, their production rates
can exceed the $3\sigma$ sensitivity of the LHC in a large part of
parameter space, so these processes may serve as a good probe of the
TC2 model.
\end{abstract}


\maketitle

\section{Introduction}

The mechanism of electroweak symmetry breaking (EWSB) remains the
most prominent mystery in elementary particle physics. Probing EWSB
will be one of the most important tasks in the high energy
colliders. Dynamical EWSB, such as technicolor (TC)
theory\cite{technicolor}, is an attractive idea that avoids the
shortcomings of triviality and unnaturalness arising from the
elementary Higgs field.

Among various kinds of technicolor theories, the topcolor scenario
\cite{topcolor} is attractive because it can explain the large top
quark mass and provides a possible EWSB mechanism. TC2 model
\cite{tc2-rev} is one of the phenomenologically viable model, which
has all essential features of the topcolor scenario. This model
predicts three CP-odd top-pions $\pi_t^0,~\pi_t^\pm$ and one CP-even
top-higgs $h_t^0$ with large couplings to the third family, which
may make these new scalar particles have a distinct experimental
signature\cite{works-tc2}. Thus, discovery of the scalar particles
in future high energy colliders would be a definite signal of new
physics beyond the standard model (SM), which would help us to
understand the scalar sector and more importantly what lies beyond
the SM.

LHC has already started its operation, and it will have considerably
capability to discover and measure almost all the quantum properties
of a SM higgs boson of any mass \cite{mh-lhc}. However, from the
theoretical view point, it would be expected that the SM is replaced
by a more fundamental theory at the TeV scale. If hadron colliders
find evidence for a new scalar state, it may not necessarily be the
SM Higgs boson. Many alternative new physics theories, such as
supersymmetry, technicolor and little Higgs, predict the existence
of new scalars or pseudo-scalar particles. These new particles may
have so large cross sections and branching fractions as to be
observable at the high energy colliders. Thus, studying the
production of the new scalars at the LHC will serve as a powerful
tool of the new physics models.


In this paper, we study how the technicolor models affect the
charged top-pion production associated with the bottom quark pair
processes $c\bar b \to \pi_t^+b\bar b$ and  $u\bar d (c\bar s) \to
\pi_t^+ b\bar b$ via the new couplings in the TC2 model. In Sec.~II,
the technicolor model relative to our calculations is briefly
reviewed.  Sec.~III shows the the numerical results for the
different processes, respectively and analysis simply the SM
backgrounds and the detectable probability. Summary and discussions
are given in Sec.~IV.

\section{About the TC2 model}
The TC2 model predicts a number of charged bosons like the top-pions
at the weak scale\cite{tc2-rev}. These scalars have large Yukawa
couplings to the quarks at tree-level, among which the top-bottom
and the charm-bottom couplings to the charged top-pion $\pi_t^\pm$
are most significant. Such couplings will induce bottom anti-bottom
pair productions associated with a charged scalar at the LHC through
the parton processes $c\bar b \to  \pi_t^+ b \bar b$ and $u\bar d
(c\bar s) \to \pi_t^+ b\bar b$. In this paper we will examine these
productions and figure out if their rates can exceed the $3\sigma$
sensitivity of the LHC. Since in the SM, such signals of the
productions have unobservably small backgrounds at the LHC, these
processes will serve as a probe for the TC2 model if their TC2
predictions can be above the $3\sigma$ sensitivity. \vspace*{0.5cm}

Before our calculations we recapitulate the basics of TC2 model. The
TC2 model\cite{tc2-rev} combines technicolor interaction with
topcolor interaction, with the former being responsible for
electroweak symmetry breaking and the latter for generating large
top quark mass. The top quark mass is generated from two sources:
one is from the extended technicolor (proportional to $\epsilon$)
and the other from the topcolor (proportional to $1-\epsilon$). So
the mass matrix of up-type quarks is composed of both extended
technicolor and topcolor contributions. The diagonalization of this
mass matrix will induce flavor changing top quark interactions in
the Yukawa couplings which involve the composite scalars
 from topcolor and technicolor condensations, respectively.

The relevant couplings with the top-pion and the fermions can be
written as\cite{FCNH}
\begin{eqnarray}
{\cal{L}}_{Y}& = &\frac{(1 - \epsilon ) m_{t}}{\sqrt{2}F_{t}}
     \frac{\sqrt{v_{w}^{2}-F_{t}^{2}}} {v_{w}} \left (
             \sqrt{2}K_{UR}^{tt *} K_{DL}^{bb}\bar{t}_R b_{L} \pi_t^-
 + \sqrt{2} K_{UR}^{tc *} K_{DL}^{bb} \bar{c}_R b_{L} \pi_t^-
 \right ) , \label{couplings}
\end{eqnarray}
where $K_{DL}$ and $K_{UR}$ are the rotation matrices that transform
the weak eigenstates of left-handed down-type and right-handed
up-type quarks to their mass eigenstates, respectively. According to
the analysis of \cite{FCNH}, their favored values are given by
\begin{equation}
 K_{DL}^{bb} \simeq 1, \hspace{5mm} K_{UR}^{tt}\simeq
\frac{m_t^\prime}{m_t} = 1-\epsilon, \hspace{5mm} K_{UR}^{tc}\leq
\sqrt{1-(K_{UR}^{tt})^2} =\sqrt{2\epsilon-\epsilon^{2}},
\label{FCSI}
\end{equation}
 In Eq.(\ref{couplings}) we neglected the mixing between
up quark and top quark. The factor $\sqrt{v_w^2-F_t^2}/v_w$ ( $v_w
\simeq 174$ GeV ) reflects the effect of the mixing between the
top-pions and the would-be Goldstone bosons \cite{9702265}.

 The total hadronic cross section for $pp \to \pi_t^+b\bar b +X$ can
 be obtained by folding the
subprocess cross section $\hat{\sigma}$ with the parton luminosity
\begin{equation}
\sigma(s)=\int_{\tau_0}^1 \!d\tau\, \frac{dL}{d\tau}\, \hat\sigma
(\hat s=s\tau) ,
\end{equation}
where $\tau_0=(2m_b+m_\pi)^2/s$, and $s$ is the $p p$ center-of-mass
energy squared. $dL/d\tau$ is the parton luminosity given by
\begin{equation}
\frac{dL}{d\tau}=\int^1_{\tau} \frac{dx}{x}[f^p_{p_1}(x,Q)
f^{p}_{p_2}(\tau/x,Q)+(p_1\leftrightarrow p_2)],\label{dis-pp}
\end{equation}
where $f^p_{p_1}$ and $f^p_{p_2}$ are the parton $p_1$ and $p_2$
distribution functions in a proton, respectively. For our case, they
could be $u$, $d$, $c$, $s$ and $b$ quark.

In our numerical calculation, the hadronic cross section at the LHC
is obtained by convoluting the parton cross section with the parton
distribution functions. In our calculations we use CTEQ6L
\cite{cteq6} to generate the parton distributions with the
renormalization scale $\mu_R $ and the factorization scale $\mu_F$
chosen to be $\mu_R = \mu_F = 2 m_b + m_\pi$.

\section{Calculations and results}

\begin{figure}[bt]
\epsfig{file=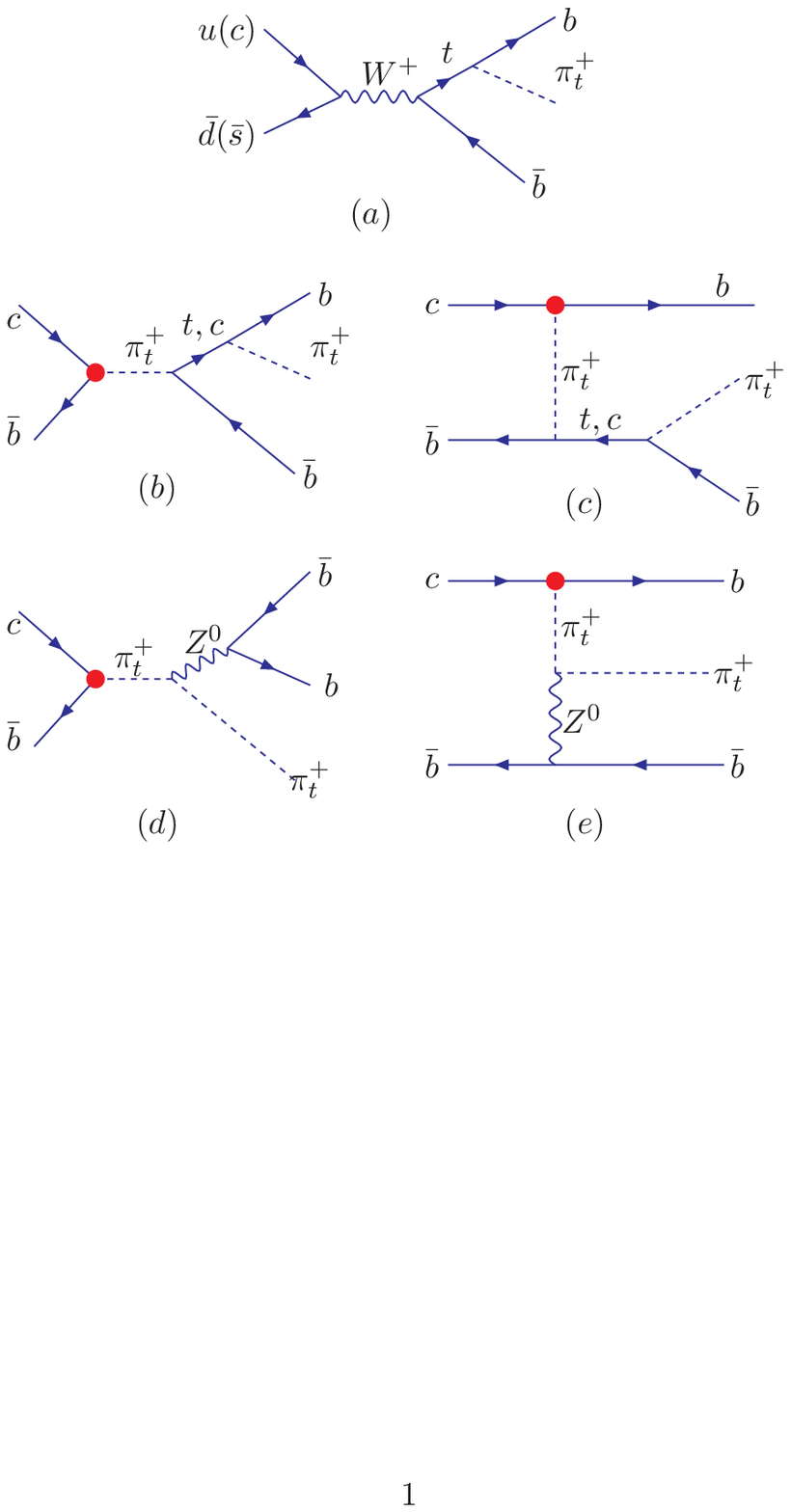,width=12cm} \vspace*{-0.5cm} \caption{Feynman
diagrams for parton-level process $u \bar d (c \bar s)\to \pi_t^+
b\bar b$ and  $c \bar b \to \pi_t^+ b\bar b$ .} \label{fig1}
\end{figure}

At the LHC, the cross sections of the charged top-pion production
comes mainly from the quark collision processes $c\bar b$, $u\bar
d$, $c\bar s \to \pi_t^+ b\bar b$, as shown in Fig.\ref{fig1}. In
our numerical calculation, we use FormCalc for the three phase space
integration \cite{formcalc}.

For the SM parameters, we will take $m_u=2$ MeV, $m_d = 3$ MeV, $m_s
= 100$ MeV, $m_c=1.27$ GeV, $m_b = 4.5$ GeV, $m_t= 172.0$ GeV, $m_Z
= 91.2 $ GeV, $m_W=80.399$ GeV \cite{pdg2010}.

The TC2 parameters involved in our calculations are the masses of
the top-pions, the parameter $K_{UR}^{tc}$, the top-pion decay
constant $F_t$ and the parameter $\epsilon$ which parametrizes the
portion of the extended-technicolor contribution to the top quark
mass. The masses of the charged top-pion mass are constrained from
the absence of $t \to \pi_t^+b$, which gives a lower bound of 165
GeV \cite{t-bpion}, and also from $R_b$ data, which yields a lower
bound of about $250$ GeV \cite{burdman}. In our numerical results we
will take $F_t=50$ GeV, $\epsilon=0.1$, $K_{UL}^{tt}=1$,
$K_{UR}^{tt}=0.9$ and retain $m_\pi$ and $K_{UR}^{tc} $ as free
parameters with $200\leq m_\pi \leq 600$ GeV and $K_{UR}^{tc} \leq
\sqrt{2\epsilon-\epsilon^{2}} = 0.43$.

 In the following we present the results for the hadronic production cross
section  $c \bar b \to \pi_t^+ b \bar b$ and $u \bar d(c \bar s) \to
\pi_t^+ b \bar b$, respectively.

\subsection{The parton level process $c\bar b \to \pi_t^+b\bar b$}

\begin{figure}[bt]
\epsfig{file=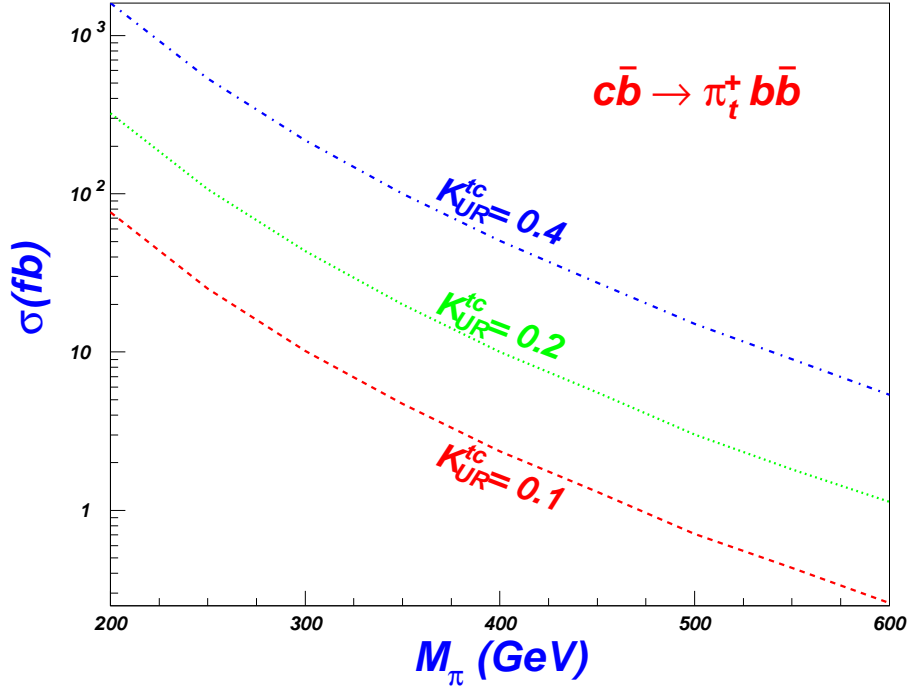,width=12cm} \vspace*{-0.5cm} \caption{Hadronic
cross section for the production via $c \bar b\to \pi_t^+ b\bar b$
at the LHC versus top-pion mass for $K_{UR}^{tc} =0.1,0.2,0.4$,
respectively.} \label{fig2}
\end{figure}
 The process is carried through out as the Fig.\ref{fig1}(b)(c)(d)(e),
 containing one or more $\pi_t^+c\bar b$ vertexes, which is proportional to the
 TC2 parameter $K_{UR}^{tc}$.

Fig.\ref{fig2} shows that the hadronic cross section versus top-pion
mass for different values of $K_{UR}^{tc}$. The cross section, which
is about several hundreds fb in most of the parameter space,
decreases with the increasing top-pion mass.

\begin{figure}[bt]
\epsfig{file=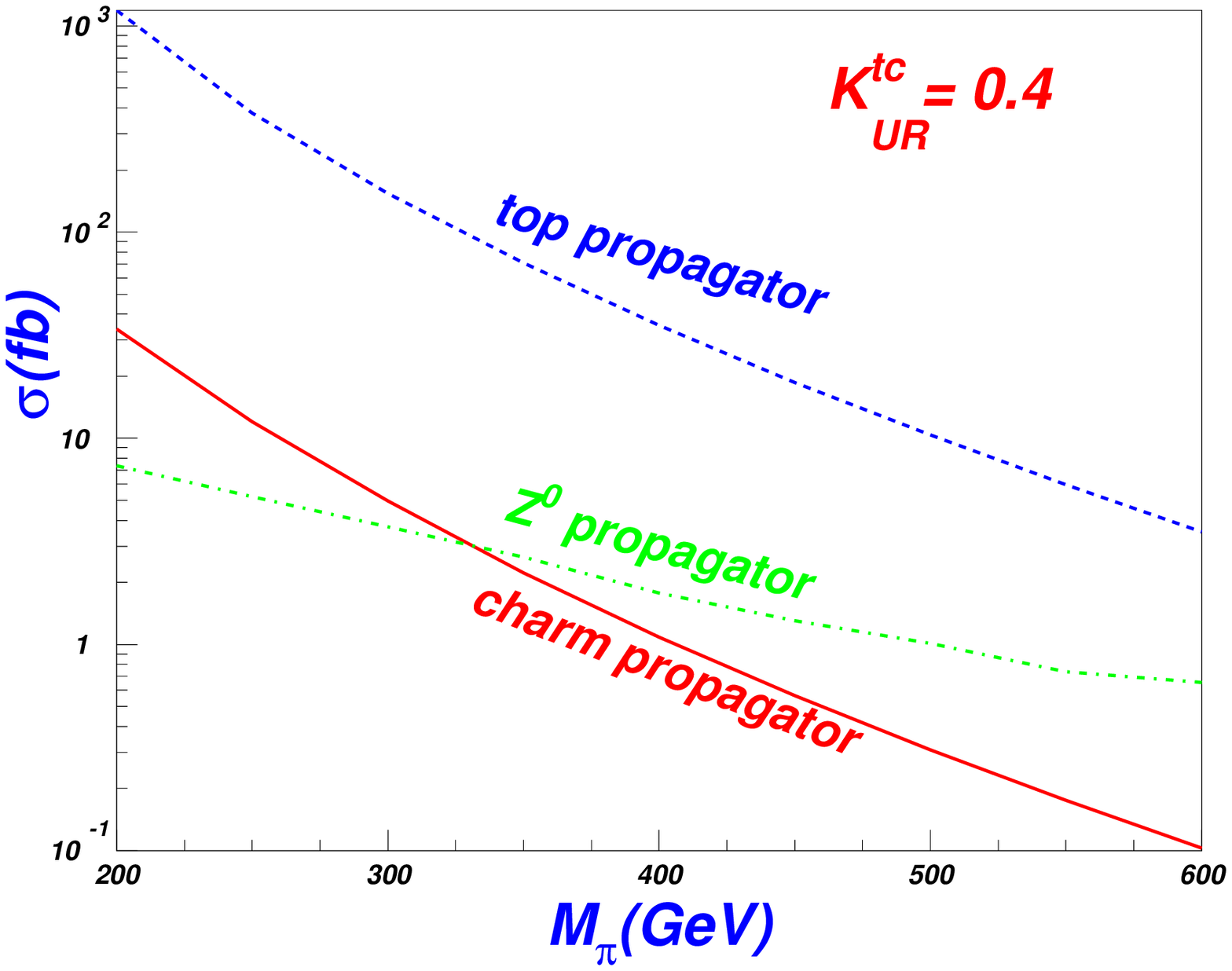,width=8cm} \vspace*{-0.5cm}
\epsfig{file=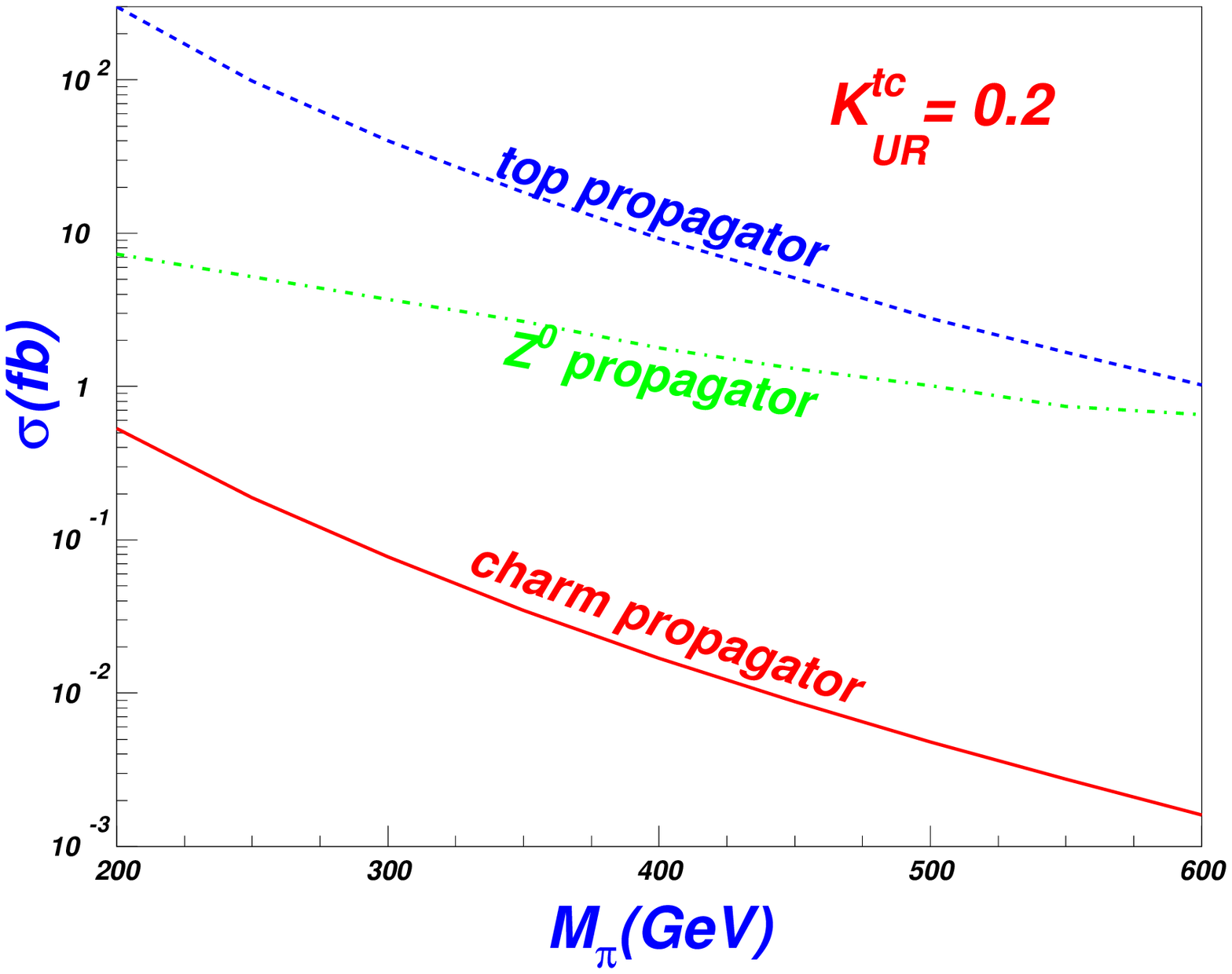,width=8cm} \caption{Different contribution to
$c \bar b\to \pi_t^+ b\bar b$ at the LHC versus top-pion mass for
$K_{UR}^{tc} =0.4$ and $0.2$, respectively.} \label{fig3}
\end{figure}

We see that the cross section increases with the increasing
$K_{UR}^{tc}$ since the cross section is mainly proportional to
$(K_{UR}^{tc})^2$, with the vertex $\pi_t^+ c\bar b$, $\sim
K_{UR}^{tc}$, which can also be seen in Fig.\ref{fig3}.

Fig.\ref{fig3} gives the different contributions of the cross
section from the exchange of top, charm quark and $Z^0$ gauge boson,
which correspond to the different contribution including top quark,
charm quark and the $Z^0$ propagator shown in
Fig.\ref{fig1}(b)(c)(d)(e). From which we can see that the top quark
propagator contribution is the most largest since $\pi_t^+ t\bar b$
couples largest.

\subsection{The processes $u\bar d \to \pi_t^+b\bar b $ and
$c\bar s\to \pi_t^+b\bar b$}

\begin{figure}[bt]
\epsfig{file=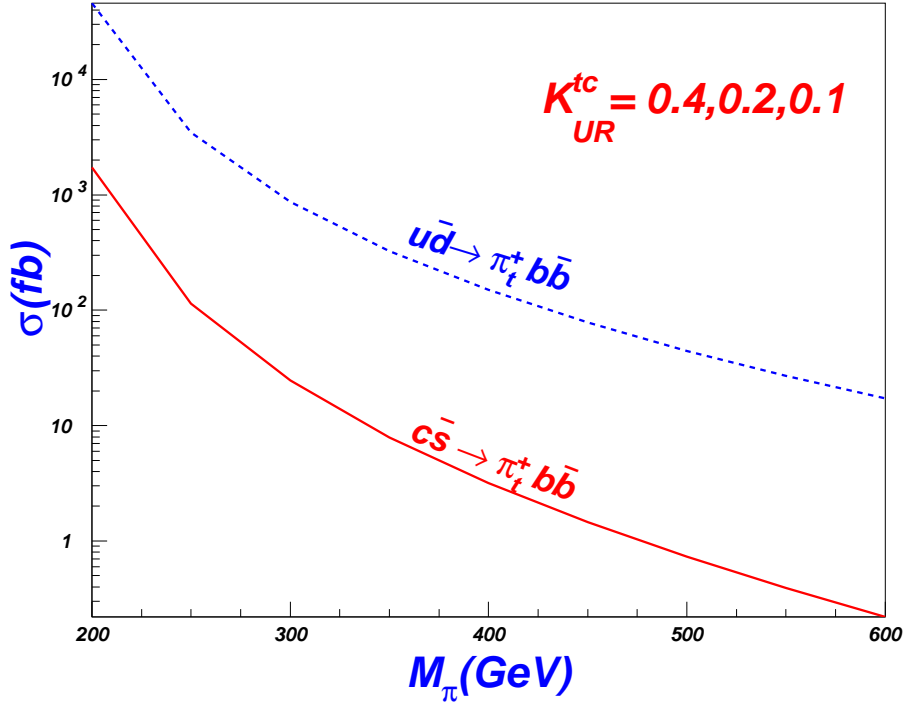,width=12cm} \vspace*{-0.5cm} \caption{Hadronic
cross section for the production via $u \bar d\to \pi_t^+ b\bar b$
and  $c \bar s\to \pi_t^+ b\bar b$ at the LHC versus top-pion mass.}
\label{fig4}
\end{figure}

Fig.\ref{fig4} shows the dependence of the cross sections of the two
processes on the top-pion masses, from which we can see that the
production rates is at the order of $1000$ fb in a large parameter
space.

We can see from Fig.\ref{fig4} that the production rate of $u\bar d
\to \pi_t^+b\bar b$ are larger than that of the $c\bar s \to
\pi_t^+b\bar b$, which is easy to understand since what makes the
difference is only the parton distribution functions when we neglect
the light quark masses.

Comparing Fig.\ref{fig2} and Fig.\ref{fig4}, we can arrive at the
conclusion that the cross sections of the $c\bar b$ collision is
smaller than that of the $u\bar d$, which is also determined mainly
by the parton distribution functions since the couplings in the two
processes are almost in the same order of the two processes, i.e,
the coupling $W^+u\bar d$ $\sim e/(\sqrt{2}\sin{\theta_W})$ is
approximately equal to that of the $\pi_t^+c\bar b$ $\sim
m_t/(\sqrt{2}F_t)K_{UR}^{tc}$, where $\theta_W$ is the Weinberg
angle. As to the process $c\bar s \to \pi_t^+b\bar b$, with the
similar parton distribution functions as that of $c\bar b \to
\pi_t^+b\bar b$, the cross sections of them are almost equivalent.

\subsection{Observability of the processes}
For the productions of $PP\to \pi_t^+b\bar b+X$ we search for the
final states from the subsequent decays $\pi_t^+\to c\bar b$,  so
the signal may be $3b+j$. The main SM background at the LHC is the
production of $ZZ$, $Zh$ and $hh$ (with the bottom quark
miss-detected as charm quark). The signal will be picked out via the
top-pion mass reconstruction since in the SM, the CKM element
$V_{cb}$ is so small that they may not produce large signal, so we
estimate the production rates arrive at $10$ fb may be detected by
the LHC.
\begin{figure}[hbt]
\epsfig{file=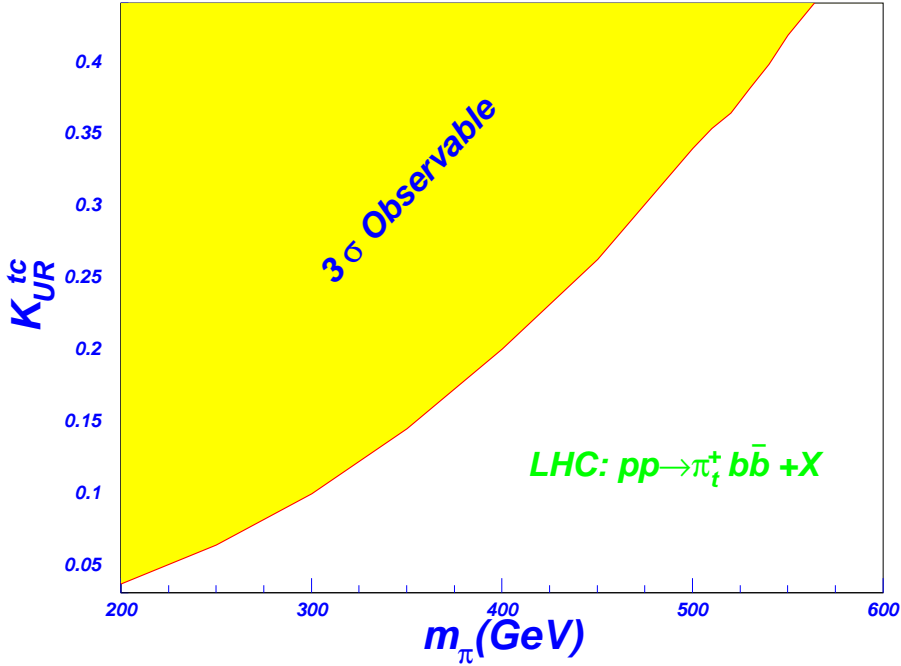,width=12cm} \vspace*{-0.5cm} \caption{The
contour of $3\sigma$ sensitivity (10 fb) for the cross section
 of the production via $c \bar b \to \pi_t^+ b\bar b$
at the LHC in the plane of $K_{UR}^{tc}$ versus top-pion mass.}
\label{contour}
\end{figure}

To show the observability of the production of $c\bar b\to \pi_t^+
b\bar b + X$, we plot in Fig.\ref{contour} the contour of the cross
section of the $3\sigma$ sensitivity ($10$ fb) in the plane of
$K_{UR}^{tc}$ versus the top-pion mass. We see that in a large part
of the parameter space the cross section can exceed the $3\sigma$
sensitivity.

For the production $u \bar d \to \pi_t^+b\bar b$, the cross section
is larger than $10$ fb in the full space of the the $m_{\pi}$(
$200\leq m_{\pi} \leq 600$ Gev). As for the production $c \bar s \to
\pi_t^+b \bar b$, as long as the top-pion mass is smaller than $340
$ GeV, the cross sections will be larger than $10$ fb and may be
detected at the LHC.

\section{Summary and conclusion}
In conclusion, we examined the charged top-pion productions
associated with a bottom pair at the LHC in topcolor-assisted
technicolor model. We found that their production rates can exceed
the $3\sigma$ sensitivity of the LHC in a large part of parameter
space. Therefore, these processes will serve as a good probe for the
topcolor-assisted technicolor model.

\section*{Acknowledgments}
We would like to thank Junjie Cao and Jin Min Yang for helpful
discussions.

\begingroup\raggedright
\endgroup

\end{document}